\documentstyle{ioplppt}
\begin{document}
\jl{3}

\date{\today}

\letter{Collective Excitations in Realistic Quantum Wires}

\author{Arne Brataas\dag\, A G Mal'shukov\ddag\, Vidar Gudmundsson\S\
  and K A Chao\dag}

\address{\dag\ Department of Physics, Norwegian University of Science
  and Technology, N-7034 Trondheim, Norway.}

\address{\ddag Institute of Spectroscopy, Russian Academy of Sciences,
  142092 Troitsk, Moscow Region, Russia.}

\address{\S\ Science Institute, University of Iceland, Dunhaga 3,
  IS-107 Reykjavik, Iceland.}

\begin{abstract}
  We have used the Hartree-Fock Random Phase Approximation (HF-RPA) to
  study the interacting electron gas in a quantum wire.  The spectra
  of intersubband spin-flip excitations reveal a considerable red
  shift with respect to single-particle HF energies.  That signals on
  appearance of collective intersubband spind-density excitations due
  to the exchange interaction.  The long wavelength dispersions of the
  intrasubband collective spin-density excitations are linear, but the
  sound velocities are renormalised due to the exchange interaction
  and screening. The in-phase intrasubband charge-density excitation
  has the long wavelength form $q[-\ln(q)]^{1/2}$. We found good
  qualitative agreement of our results with experimental observations.
\end{abstract}


A semiconductor quantum wire can be fabricated by applying a voltage
with a microstructured gate to a 2D electron gas. The single-particle
energy (SPE) spectrum typically consists of subbands separated by
several $meV$. At a 1D electron density about $10^6$cm$^{-1}$ more
than one subband can be occupied.  In recent years progress has been
made in spectroscopic study of such systems. In angular resolved Raman
spectra of GaAs quantum
wires\cite{Egeler90:1804,Goni91:3298,Schmeller94:14778}, collective
spin-density excitations (SDE) and charge-density excitations (CDE, or
plasmons) were observed. The measured spectra cover from low to high
frequency, and for low-frequency intrasubband excitations, the
wave-vector dependence of the spin-wave energy was found to be linear.
The correct interpretation of these experiments allows us not only to
understand the interesting physical processes, but also to access
important physical parameters.

The exchange interaction is crucial to the collective intrasubband and
intersubband SDE. Similar to the direct long range Coulomb interaction
which leads to the depolarisation shift of single-particle excitations
and to the appearance of collective plasma modes, exchange interaction
gives rise to the red-shift of SPE. If the red-shift is sufficiently
large, the collective SDE with a sizable oscillator strength splits
off the continuum of SPE. For semiconductor quantum wells, such
split-off appears in the Hartree-Fock Random Phase Approximation
(HF-RPA)\cite{Ando82:3893,Katayama84:1615,Tselis84:3318,Eliasson87:5569}.
Thus, it is important to perform a HF-RPA analysis on collective
electron excitations in a realistic GaAs quantum wire with full
Coulomb interaction, and compare the results with measured
spectra\cite{Goni91:3298,Schmeller94:14778}. The selfconsistent and
conserving\cite{Baym61:287} HF-RPA is suitable for this task, because
we will calculate the two-particle spectra but not single-particle
properties. As in using any approximation, the HF-RPA calculation also
contains error. However, our HF-RPA results agree very well with
experimental measurements.

In this Letter we will first outline the HF-RPA method to express the
spin or charge correlation functions in terms of the corresponding
spin- or charge-density induced matrix, which satisfies an eigenvalue
matrix equation. In a realistic quantum wire, for high energy
intrasubband excitations and for excitation spectra when more than one
subband is occupied, the self-consistent equations have to be solved
numerically. However, for the case that only one subband is occupied,
the analytical expressions of SDE and CDE are derived. With additional
approximation applied to the HF-RPA analytical solutions, they reduce
to known results.  This may be coincidental, but is outside the
central theme of this Letter.

To define the quantum wire, let us start from an electron gas in a
narrow quantum well with interfaces parallel to the $x$-$y$ plane, and
the well width shorter than any other relevant length scales. We
consider the realistic experimental situation that only the lowest
subband in the quantum well is occupied, so the motion of electrons in
the $z$ direction can be ignored. By applying a properly designed gate
potential, a quantum wire along the $x$ axis is fabricated with a
longitudinal constant electrostatic potential, but a transverse
parabolic one $V_{\text{c}}(y)=m^*\omega_0^2y^2$/2 along the $y$ axis,
where $m^*$ is the electronic effective mass. The quantum wire has a
finite length $L_x$, and we assume periodic boundary conditions. In
the absence of electron-electron interaction, the single particle
eigenfunctions are simply $\psi_{nk}(x,y)$=$L_x^{-1/2}
e^{ikx}\psi_{nk}(y)$, with the corresponding eigenenergies
$E_{nk}$=$\hbar\omega_0 (n$+$1/2)$+$(\hbar^2 k^2)/(2m^*)$, where $n$
is the subband index, $k$=$integer\times 2\pi/L_x$, and $\psi_{nk}(y)$
the harmonic oscillator wave function. The transverse confinement
length of the electrons is $l_0$=$[\hbar /( m^*\omega_0)]^{1/2}$.

When the electron-electron interaction is turned on, we will use the
HF approximation
\begin{eqnarray}
  \{ & - &\frac{\hbar^2}{2 m^*} \nabla^2 + V_{\text{c}}(y) +
  \frac{2e^2}{\kappa} \int d\vec{r'} \sum_b f_b
  \frac{|\psi_b(\vec{r'})|^2}{|\vec{r}-\vec{r'}|} \} \psi_a(\vec{r})
  \nonumber \\ & - & \frac{e^2}{\kappa} \int d\vec{r'} \sum_b f_b
  \frac{\psi_b^*(\vec{r'})\psi_b(\vec{r})}
  {|\vec{r}-\vec{r'}|}\psi_a(\vec{r'}) = \epsilon_a^{HF}
  \psi_a(\vec{r})
        \label{HFground}
\end{eqnarray}
to derive a complete orthonormal basis $\{\psi_a(\vec{r})\}$ of
quasi-particle Hartree-Fock states, where $\epsilon_a^{HF}$ is the
Hartree-Fock energy, $f_a$ is the Fermi occupation factor and $\kappa$
the dielectric constant of the surrounding medium. In terms of this
basis set, we will use the corresponding time dependent HF (HF-RPA) to
calculate the charge-density and the spin-density correlation
functions, which describe the self-consistent linear response of the
electron gas to an external perturbation. In this Letter, we restrict
ourselves to the nonmagnetic ground state.

For inelastic light scattering not close to the band gap resonance,
the Raman intensities in polarised and depolarised scattering
geometries are proportional, respectively, to the imaginary parts of
the charge-charge correlation function and spin-spin correlation
function\cite{Hamilton69}. If we define $\hat{\rho}(\vec{q},t)$ as the
Fourier transform of the charge-density operator and
$\hat{\sigma}(\vec{q},t)$ the Fourier transform of the spin-density
operator along the spin quantisation axis, then, these correlation
functions are $\chi^{\nu}(\vec{q},\omega) = - \frac{i}{\hbar}
\int_{0}^{\infty} dt e^{i \omega t} <|\left[\hat{\nu}(\vec{q},t),
  \hat{\nu}(-\vec{q},0) \right] |> $ with $\nu$=$\sigma,\rho$, where
$\langle | \ldots |\rangle$ denotes a thermodynamic average. In
HF-RPA, they can be expressed as $\chi^{\nu}(\vec{q},\omega) =
\sum_{cd}K_{cd}^{\nu}(\vec{q},\omega) \langle c | \exp{- i \vec{q}
  \cdot \vec{r}} | d\rangle $ in terms of the $\nu$-density induced
matrix $K^{\nu}$. The charge-density induced matrix satisfies
\begin{eqnarray}
  K_{ab}^{\rho}(\vec{q},\omega) & = & \frac{f_{b}-f_{a}}{\hbar \omega
    + i 0^+- (\epsilon_{a}^{HF}-\epsilon_{b}^{HF})} [ 2\langle
  a|e^{-i\vec{q}\cdot\vec{r}}|b\rangle^* \nonumber \\ & - & \sum_{cd}
  (2V_{cb;ad} - V_{bc;ad}) K_{cd}^{\rho}(\vec{q},\omega) ] \, ,
\label{indcharge}
\end{eqnarray}
where $V_{ab;cd}$=$\int d\vec{r}\int d\vec{r'}\psi_a^*(\vec{r})
\psi_d(\vec{r})\psi_b^*(\vec{r'})\psi_c(\vec{r'})/|\vec{r}-\vec{r'}|$
is the Coulomb matrix element in the basis of the HF single-particle
states. When $\nu$=$\sigma$, there is no direct Coulomb interaction
(Hartree term) between the excited spins. If the exchange term (Fock
term) is neglected, the spin-density excitation spectra are simply
given by the quasi-particle Hartree-Fock energies.  In HF-RPA, the
spin-density excitations are shifted from the quasi-particle
Hartree-Fock energies due to the exchange terms. As a result,
collective spin-density excitation may appear in the spectra.  This
shift is a measure of the effective strength of the exchange term in
the system.

We will use the above HF-RPA analysis to investigate collective
excitations in a multisubband system. This has to be done numerically,
and the first step is to derive the quasiparticle basis set
$\{\psi_a(\vec{r})\}$ by solving (\ref{HFground}) self-consistently
via iteration. The functional basis must be sufficient large in order
to ensure the required accuracy for the quasiparticle energy spectra
and the spatial electron density of the ground state. When using this
self-consistent quasiparticle basis set to calculate collective
excitations, the required accuracy for the peak position of an
excitation energy is the experimental linewidth $\simeq$0.1 $meV$.

The intrinsic materials parameters of a GaAs quantum wire required in
our calculation are $m^*$=$0.067m_0$ and $\kappa$=12.4, where $m_0$ is
the electron mass. The extrinsic sample parameters are the total
electron density $n$, the transverse potential $V_c(y)$, and the
length of the quantum wire $L_x$. Our calculations are based on a
sample with $n$=10.4$\times$$10^5$ $cm^{-1}$, which has been
investigated experimentally in Ref.~\cite{Schmeller94:14778}.  The
transverse potential $V_c(y)$ is determined by the experimental
conditions. Since the experimental sample does not have a parabolic
potential as can be seen from the intersubband CDE, we will adjust
$\hbar\omega_0$ to fit the subband electron occupations. When we set
$\hbar\omega_0$=7.9 $meV$ ($l_0$=120 {\AA}), the derived
self-consistent Hartree-Fock subband spacing between the two lowest
subbands is 5.4 $meV$ at the zone center and 5.6 $meV$ at the Fermi
level. The electron densities in the two occupied subbands are
$n_0$=6.4$\times$$10^5$ $cm^{-1}$ for the lowest subband (labeled by
$i$=0), and $n_1$=4.0$\times$$10^5$ $cm^{-1}$ for the second lowest
subband (labeled by $i$=1).  The physical properties of the quantum
wire are insensitive to its length, provided that the wire is
sufficiently long. Therefore, we set $L_x$=1.0 $\mu$m. The
experiment\cite{Schmeller94:14778} was done at a temperature 1.7 K,
which is the temperature used in our calculation.

Let $k_i$ and $v_i$ be, respectively, the Fermi wave-vector and the
Fermi velocity of the $i$th subband. In the limit of long wavelength
$q$$\rightarrow$0, the transverse excitation is forbidden (or allowed)
if the excitation energy is low (or high), and so the corresponding
physical processes are dominated by intrasubband (or intersubband)
transitions.  When $q$ increases to $k_0$-$k_1$, the low energy
intersubband transitions are activated. However, collective
excitations with large energy and/or large $q$ are strongly damped. To
illustrate the main features of long wavelength excitations, let us
consider a very simply case by neglecting the electron self-energies
and the vertex correction. In this case we obtain analytical results
for both the SDE and CDE energies. The SDE dispersion is linear in
$q$, with sound velocities $v_0$ and $v_1$ for intrasubband
excitations in the two lowest subbands. With two subbands occupied,
the CDE dispersion has two branches. The in-phase mode is
\begin{equation}
  \omega_{\rho}^+(q) = |q| \sqrt{2 (v_0 + v_1) V(q)/\hbar\pi } \, ,
\label{2charge+}
\end{equation}
where $V(q)$ is the Fourier transform of the Coulomb potential
$(e^2/\kappa)[(x$-$x')^2$+$(y$-$y')^2]^{-1/2}$, and the out-of-phase
mode is
\begin{equation}
  \omega_{\rho}^-(q) = |q| \sqrt{v_{0}v_{1}} \, .
\label{2charge-}
\end{equation}
Such two-band SDE and CDE dispersions are exactly the same results as
Schulz\cite{Schulz93:1864} obtained for a two-subband
Tomonaga-Luttinger model (TLM)\cite{Haldane81:2585}.

When we turn on the intrasubband and intersubband exchange
interaction, as well as the screening, the SDE and CDE energies will
be modified from the above expressions, and have to be derived
numerically.  In the polarised spectrum, the higher energy region is
dominated by strong intersubband CDE. At $q$=0, the calculated CDE
spectrum consists of a single peak which is characteristic to
transverse parabolic confinement potentials\cite{Kohn61:1242}. In the
region of finite $q$ the intersubband HF SPE energies show up at 5.5
$meV$.  In the spin-flip depolarised Raman spectrum, we obtained the
red shift of intersubband SDE with respect to HF SPE. The SDE has a
dominant weight of the spectrum and appears at a resonance energy of
2.1 $meV$, which is shifted from the HF SPE, indicating the importance
of vertex correction. These overall features of our numerical results
have been observed experimentally\cite{Goni91:3298,Schmeller94:14778}.

In Fig.~\ref{CDESDE} we show the calculated intrasubband SDE and CDE
in the region of energy and frequency for which accurate experimental
data are available.  The mode with strong intensity is labeled as
SDE0, and the next mode with weaker intensity is labeled as SDE1. The
inset gives their dispersions, which are linear for small $q$, with
the corresponding sound velocities $v_{\sigma,0}$=0.9$v_0$ and
$v_{\sigma,1}$=0.7$v_1$. These sound velocities are smaller than the
respective Fermi velocities because of the exchange screening of SDE
induced by the intersubband virtual transitions, and the
renormalisation due to the intra- and intersubband exchange
interaction. The experiments\cite{Goni91:3298,Schmeller94:14778} seem
to have detected only $v_{\sigma,0}$, which corresponds to the high
energy SDE near the Fermi wave vector $k_0$ of the lowest subband. It
is probably the low electron density in the second lowest suband that
causes the velocity $v_{\sigma,1}$ of the SDE around $k_1$ to be too
small to be observed experimentally. The spin velocity $v_{\sigma,0}$
agrees quantitatively with the experiment\cite{Schmeller94:14778}.
For sufficiently large $q$, because of the finite length of our
quantum wire, the Landau damping shows up in Fig.~\ref{CDESDE} in the
form of enhanced intensities of satellite single-particle peaks around
the SDE, instead of a broad band if the quantum wire is infinitely
long.  The mode SDE1 is not damped, since it does not enter the region
of SPEs.

The dispersions of the CDE and SDE are also plotted in the inset in
Fig.~\ref{CDESDE}. In the long wavelength limit, the in-phase CDE+ can
be well fitted with $q[-\ln(q)]^{-1/2}$ according to (\ref{2charge+}),
as expected for 1D electron gas\cite{Schulz93:1864,Li89:5860}. The
out-of-phase CDE- mode, corresponding to (\ref{2charge-}) with an
expected linear dispersion\cite{Schulz93:1864}, appears in our
calculation as a weak band with a sound velocity $1.2(v_0v_1)^{1/2}$.
This happens to be the same as the sound velocity of the SDE0, an
accidental result for this specific sample.  The CDE- mode has a much
lower intensity compared to the CDE+ mode.  Checking against the
experiment\cite{Goni91:3298}, we believe the the lower energy band
observed in the low frequency polarised Raman spectrum is our
calculated CDE- mode. The decay of the plasmon (CDE+) and the CDE- at
higher $q$ due to the Landau damping is stronger than the decay of
SDE0 and SDE1, as shown in Fig.~\ref{CDESDE}.

After the above complete study of a realistic quantum wire, perhaps it
is worthwhile to mention some surprising findings for the special case
that only the lowest subband occupied at zero temperature, namely, the
pure 1D system including spin degrees of freedom. In this case we can
drop the band index. For SDE, the eigenvalue equation has the form
\begin{equation}
  K^{\sigma}_{k}(q,\omega) = - \frac{f_{k-q/2} - f_{k+q/2}} {\hbar
    \omega - (\epsilon_{k+q/2}^{HF}-\epsilon_{k-q/2}^{HF})}
  \frac{1}{2\pi}\int dk^{\prime} V(k-k^{\prime})
  K^{\sigma}_{k^{\prime}}(q,\omega) \, ,
\label{SDEone}
\end{equation}
where $V(k)$=$(2e^2/\kappa)K_0(q l_0)$ for the quasi-1D Coulomb
potential $(e^2/\kappa)[(x$-$x')^2$+$l_0^2]^{-1/2}$, which has a
logarithmic singularity $(-2e^2/\kappa)\ln{ql_0}$ as
$q$$\rightarrow$0. Since the Hartree self-energy is canceled by the
potential energy due to the positive background charges, the
difference in exchange self-energy at the Fermi energy,
$\Delta_x(q)$=$\Sigma_x(k_0$+$q/2)$-$\Sigma_x(k_0$-$q/2)$ is $
\Delta_x(q) = \frac{1}{2 \pi} \int_{-q/2}^{q/2}dk^{\prime} \left[
  V(k^{\prime}) - V(k^{\prime}-2k_0) \right] $.
We define $y(k)$ by
$K_k^{\sigma}$$\equiv$$(f_{k-q/2}$-$f_{k+q/2})y(k)$, and at $k$=$k_0$
rewrite (\ref{SDEone}) as
\begin{eqnarray}
  && \int_{-q/2}^{q/2} d k^{\prime} \left[
    -V(k^{\prime})y(k_0+k^{\prime}) +
    V(k^{\prime}-2k_0)y(-k_0+k^{\prime}) \right] \nonumber \\ && =
  2\pi\left[ \hbar\omega -\hbar qv_0 - \Delta_x(q)\right] y(k_0) \, .
\label{add1}
\end{eqnarray}
In the limit of long wavelength, by substituting
$y(k_0+k^{\prime})$$\simeq$$y(k_0)$ and
$y(-k_0+k^{\prime})$$\simeq$$y(-k_0)$ into the above equation, we see
that the contributions of Coulomb potential in the exchange energy and
the vertex corrections, both behave as $q\ln(q)$, exactly cancelling
each other.  Hence, the $q\ln(q)$ behaviour of low-energy
single-particle excitation due to the exchange self-energy is removed
from the SDE and the plasmon energies, and so only the exchange terms
at momentum $2k_0$ are important.  An equation similar to (\ref{add1})
for $y_k$ at $k=-k_0$ can also be derived. In the long wavelength
limit, we find the SDE energy
\begin{equation}
  \omega_{\sigma}(q) = |q| v_0\sqrt{1 - 2g_0} \, ,
\label{1spin}
\end{equation}
where $g_0$=$V(2k_0)/2\hbar\pi v_0$.  The sound velocity is reduced
with respect to the bare Fermi velocity. We have also solved for the
CDE and found the plasmon dispersion
\begin{equation}
\omega_{\rho}(q) = |q|v_0 \sqrt{
2V(g)/\hbar\pi + 1 - 2g_0} \, .
\label{1charge}
\end{equation}
For a long-range Coulomb potential $V(q)$$\sim$$\ln{q}$, the plasmon
energy is not affected by the exchange interaction within the HF-RPA.
These results (\ref{1spin}) and (\ref{1charge}), derived with full
Coulomb interaction, differ from those obtained by
Schulz\cite{Schulz93:1864} for the TLM including a nonsingular
backscattering matrix element at $2k_0$, which mixes the right- and
the left-traveling modes. Nevertheless, in (\ref{SDEone}), if we
replace the Hartree-Fock energy by the bare single-particle energy,
and omit the singular part around V(q=0) in the vertex correction,
then our results exactly reduce to Schulz's results for the SDE and
the plasmon energies. Whether this finding, as well as
(\ref{2charge+}) and (\ref{2charge-}), is simply a coincidence remains
to be clarified.

To close this Letter, we should mention that according to our
analysis, the frequency of the lowest energy SDE decreases as $q$
increases towards $k_1$, and then vanishes at a value of $q$ close to
$2k_1$. This behaviour suggests a low temperature intrinsic
instability of the electron gas against the formation of spin-density
waves (SDW), namely, the Peierls instability. When this instability
emerges from our mean-field analysis, we must take into account the
SDW long-range order in the ground state.  On the other hand, the
results of the TLM with backscattering predict the absence of SDW
long-range order, but instead the appearance of a slowly decaying
Wigner crystal at $4k_F$\cite{Schulz93:1864}.  We guess that the TLM
with backscattering describes the real ground state better than
HF-RPA.  Hence, we believe that our analysis based on a ground state
without SDW long range order is a better approximation than that
including SDW in the ground state.

A. B. would like to thank A.~Sudb\o\ and L.~J.~Sham for stimulating
discussions.  This research was supported in part by a NorFa Grant.

\Figures

\begin{figure}
\caption{
  Intensity of intrasubband SDE and CDE as a function of the frequency
  shift for different wave vectors $q$.  The upper part shows the
  dispersion of intrasubband CDE and intrasubband SDE.  The quantum
  wire is specified in the text.}
\label{CDESDE}
\end{figure}

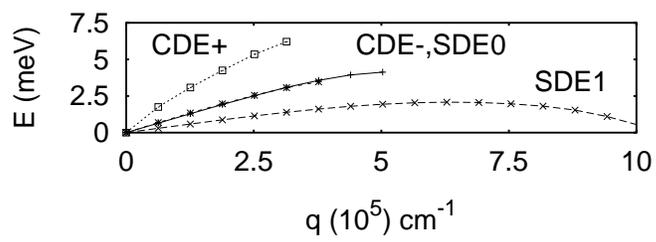
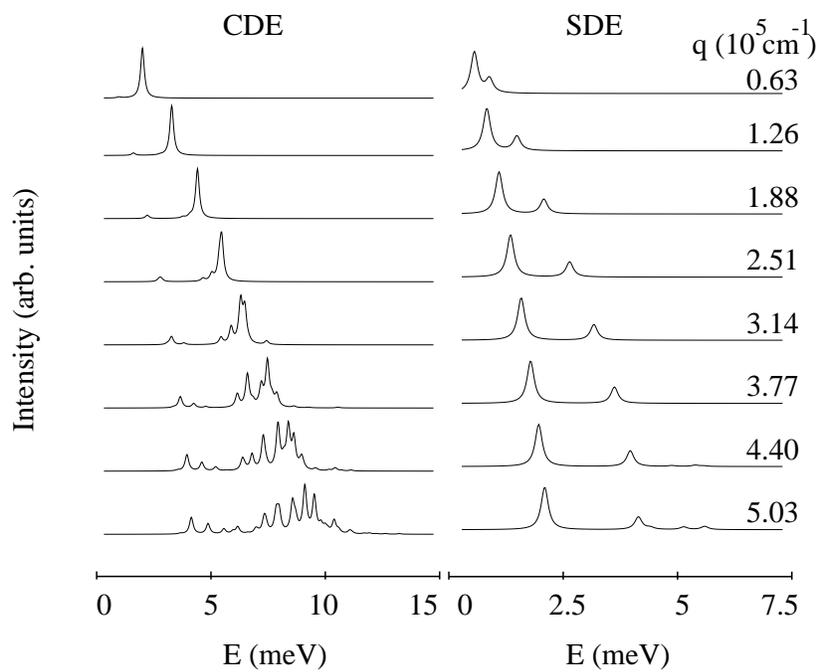
\end{document}